\documentclass[prd,aps,preprint,tightenlines,nofootinbib,superscriptaddress]{revtex4}
\usepackage{epsfig}
\usepackage{amsmath}
\input{epsf}
\usepackage{psfrag}
\usepackage{xcolor}

\begin{document}


\vspace*{2cm}
\title{On  the  kinematic constraint in BFKL near threshold region}

\author{Hao-yu  Liu}
 \affiliation{\normalsize\it Center of Advanced Quantum Studies, Department of Physics,
Beijing Normal University, Beijing, China}

\author{Xiao-hui Liu}
 \affiliation{\normalsize\it Center of Advanced Quantum Studies, Department of Physics,
Beijing Normal University, Beijing, China}


\author{Yu Shi}
 \affiliation{\normalsize\it Key Laboratory of
Particle Physics and Particle Irradiation (MOE),Institute of
Frontier and Interdisciplinary Science, Shandong University,
QingDao, China }

\author{Du-xin Zheng}
 \affiliation{\normalsize\it School of Physics and State Key Laboratory of Nuclear Physics and Technology, Peking University, Beijing, China}

\author{Jian Zhou}
 \affiliation{\normalsize\it Key Laboratory of
Particle Physics and Particle Irradiation (MOE),Institute of
Frontier and Interdisciplinary Science, Shandong University,
QingDao, China }

\begin{abstract}
We introduce a modified Balitskii-Fadin-Kuraev-Lipatov (BFKL) equation with  the rapidity veto that originates from external kinematic constraint. Though it is a formally sub-leading power contribution, such kinematic effect becomes very important especially in the threshold region, where there is no sufficient phase space for the small $x$ evolution to be fully developed. We also investigate the phenomenological consequences of the kinematic constraint BFKL equation in the forward particle  production processes in pp and eA collisions.  It is shown that particle yield at large transverse momentum is significantly suppressed in the threshold region.

\end{abstract}

\maketitle

\title{Forward particle production in pA collisions}

\author{....
 \\[0.3cm]
{\normalsize\it School of Physics and Key Laboratory of Particle Physics and Particle Irradiation (MOE),} \\
  {\normalsize\it Shandong University, QingDao, Shandong, 266237, China}
}

\date{\today}

\section{Introduction} \label{s:intro}
Understanding the behavior of parton densities at small $x$ is one of the important aspects of the nucleon/nucleus internal structure studies.  It is also crucial for describing the experimental data of high energy scattering processes in ep/eA and pp/pA collisions. The small $x$  evolution of the unintegrated gluon distribution is governed by the famous BFKL equation~\cite{Kuraev:1977fs,Balitsky:1978ic} in the dilute limit. When the parton density is so high  that the gluon fusion process becomes important, the proper evolution equation for describing small $x$ dynamics is the JIMWLK equation~\cite{Jalilian-Marian:1997jhx,Jalilian-Marian:1997qno,Jalilian-Marian:1996mkd,Iancu:2001md} or its large $N_c$ version: the BK equation~\cite{Balitsky:1995ub,Kovchegov:1999yj}. To improve the precision of the BFKL/BK predication for phenomenology, it is necessary to carry out the program for computing the next-to-leading-order (NLO) corrections to the small $x$ evolution equations, which has been achieved in the recent years~\cite{Fadin:1996nw,Ciafaloni:1998gs,Balitsky:2007feb,Balitsky:2013fea}. Another major development along the direction is the establishment of the joint small $x$ and $k_\perp$ resummation formalism~\cite{Mueller:2012uf,Mueller:2013wwa,Zhou:2016tfe,Xiao:2017yya,Zhou:2018lfq}. On the other hand, the calculations of the impact factors in various high energy scattering processes have been pushed to the NLO level as well~\cite{Chirilli:2011km,Chirilli:2012jd,Balitsky:2012bs,Beuf:2011xd,Boussarie:2016ogo,Boussarie:2016bkq,Beuf:2017bpd,Hanninen:2017ddy,Roy:2018jxq,Roy:2019hwr,Boussarie:2019ero,Mantysaari:2021ryb,Iancu:2020mos,Beuf:2021qqa,Caucal:2021ent}.

A known issue with the NLO BFKL or BK equations is that the next-to-leading logarithmic (NLL) contribution are large compared to the leading log $\ln \frac{1}{x}$, which renders the result unstable. These sub-leading logs  thus need to be resummed to all orders. One way of including part of the NLL contributions is to enforce the kinematic constraint in the small $x$ evolution equations~\cite{Ciafaloni:1987ur,Catani:1989yc,Andersson:1995ju,Kwiecinski:1996td,Kwiecinski:1997ee,Smallx:2006lvi}. The implementation of the kinematic constraint was motivated  by the  requirement that the off-shellness of the exchanged gluon in the BFKL cascade are dominated by the transverse components.  Though such kinematic corrections formally are the next-to-leading logarithmic contribution, they are shown to be numerically very large as compared to the leading log result. It thus has long been recognized as the necessary ingredients for the phenomenological application of small $x$ resummation formalism~\cite{Kwiecinski:1997ee,Deak:2019wms}.  Other elaborated resummation schemes can be found in Refs.~\cite{Iancu:2015vea,Ducloue:2019ezk,Zheng:2019zul,Xiang:2020xxe,Xiang:2021rcy}.

In this work, we introduce a novel NLL BFKL equation that originates from the external kinematic constraint. The key observation that motivates this modified BFKL equation is that the minimal  longitudinal momentum fraction of the radiated gluon $\Delta x$ is finite.   The minimal $\Delta x$ is determined by the onshell condition $\Delta x P^+ > \frac{l_\perp^2}{2P^-_\text{max}}$ where $l_\perp$ is the radiated gluon's transverse momentum and $P^-_\text{max}$ is the largest possible minus component of longitudinal momentum transferred from the projectile.  By imposing this kinematic constraint, the resulting small $x$ log is $\int_{x_g+\Delta x}^1 \frac{dx}{x}=\ln \frac{1}{x_g+\Delta x}$ instead of the conventional one $\ln \frac{1}{x_g}$. Obviously, such kinematic constraint BFKL will slow down small $x$ evolution as compared to the standard BFKL.  And in particular, this effect is greatly enhanced  in the threshold region  where the available $P^-_\text{max}$ is very small. The main goal of the paper is to  numerically solve the kinematic constraint BFKL equation, and investigate its impact on the phenomenology studies.

The paper is structured as follows. In Sec.II, we first discuss  the BFKL equation with kinematic constraint and present numerical solutions.  In Sec.III, we study the inclusive forward hadron production in pp collisions by applying the kinematic constraint BFKL evolution. It is shown that subtracting the NLL BFKL instead of the leading log BFKL from the hard part leads  to a more stable NLO result.  We also make predictions for forward jet production in the small $x$ limit in SIDIS process. It is found that transverse momentum spectrum of the cross section is  suppressed due to the limited phase space available for real radiations near threshold region.   The paper is summarized in Sec. IV.

\section{The BFKL equation with external kinematic constraint}
We start the discussion about the kinematic constraint BFKL with an explicit example: inclusive forward particle production in pp collisions.  The leading order calculation of this process  is formulated in a hybrid approach where the projectile is described by the normal parton distribution function(PDF), while the  target is treated  in the small $x$ formalism~\cite{Dumitru:2001jn}. If the outgoing parton carries longitudinal momentum $x_p P^-$, the maximal minus component of the exchanged gluon's momentum is  $(1-x_p)P^-$  according to the four momentum conservation. When $x_p$ approaches  1, the small $x$ evolution  is hampered due to the limited phase space for the real correction to the BFKL kernel, particularly at high transverse momentum.  Note that such kinematic constraint  is different from the one studied in the literature~\cite{Ciafaloni:1987ur,Catani:1989yc,Andersson:1995ju,Kwiecinski:1996td,Kwiecinski:1997ee}. 

\begin{figure}[htbp]
\includegraphics[scale=0.6]{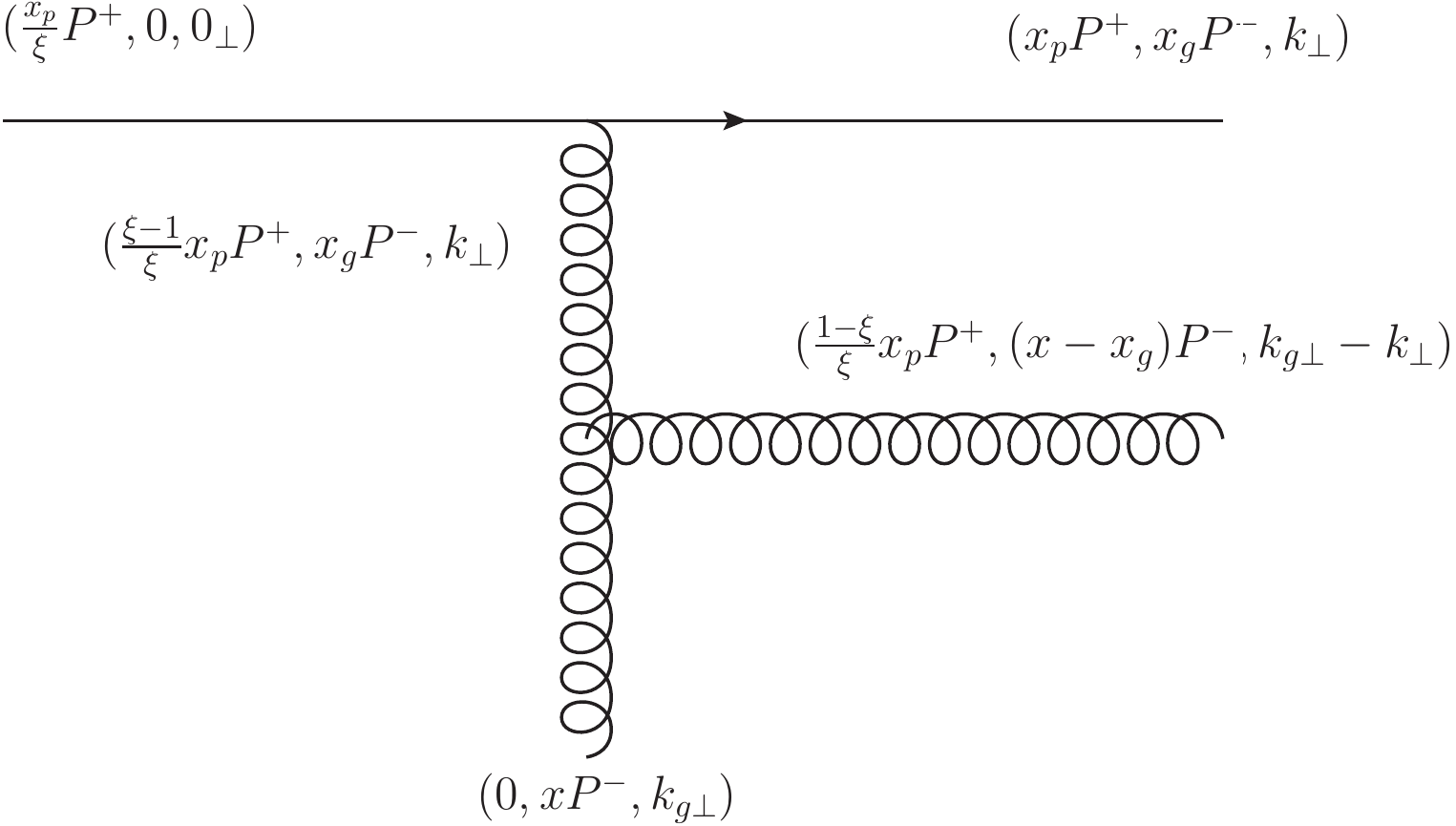}
\caption{ A  real diagram representing gluon emission in the BFKL chain  in the forward particle production from quark channel in pp  collisions. The incident quark moves along z direction, while the target moves in the opposite direction. The light-cone momenta variables are commonly defined. } 
\label{illustration}
\end{figure}

We now explain how to  implement  this new type kinemaitical constraint in the BFKL evolution.  In the leading logarithmic approximation, the standard BFKL equation for the unpolarized gluon TMD(often referred to as the unintegrated gluon distribution as well) reads,
\begin{eqnarray}
  G(x_g,k_{\perp}) \!\!&=&\!\! G_0(x_g,k_{\perp})
   \\ &+&\!\! \frac{ \alpha_s N_c}{\pi^2}  \!\! \int_{x_g}^{1}\! \frac{d x}{x} \int \!\! d^2 k_{g\perp}   \! \left \{  \frac{ G(x,k_{g\perp})}{(k_\perp-k_{g\perp})^2} 
 -  \frac{k_\perp^2 G(x,k_{\perp})}{(k_\perp-k_{g\perp})^2 \left [k_{g\perp}^2+(k_\perp-k_{g\perp})^2\right ]  
 }   \right \} \nonumber
 \label{sBFKL}
 \end{eqnarray}
 where $G_0(x_g,k_{\perp})$ is the bare gluon TMD. $k_\perp$ and $k_\perp-k_{g\perp}$ are the transverse momenta of the exchanged gluon and the emitted gluon respectively.    The flow of the momenta in the BFKL cascade is illustrated in Fig.~\ref{illustration}.  To isolate the leading log contribution in the each rung of the BFKL ladder diagram, the virtuality of the exchanged gluon is required to be  
\begin{eqnarray} 
| 2\frac{\xi-1}{\xi}x_g x_p P^+ P^- -k_\perp^2| \sim k_\perp^2
\end{eqnarray}
 This subsequently converts  to the condition,
 \begin{eqnarray}
 2 \frac{1-\xi}{\xi}x_g x_p P^+ P^- < c k_\perp^2 
 \label{uneq}
\end{eqnarray}
where we set the coefficient $c$ to be 1 for simplicity.
 The on-shell condition for the emitted gluon yields,
 \begin{eqnarray}
 2 \frac{1-\xi}{\xi}(x-x_g) x_p P^+ P^-=(k_{g\perp}-k_\perp)^2  
 \label{onshell}
\end{eqnarray}
Moreover, the strong ordering of the longitudinal momenta in the BFKL kinematics implies, 
 \begin{eqnarray}
  x_g \ll x
\end{eqnarray}
Replacing $x-x_g$ with $x $ in Eq.~(\ref{onshell}) and inserting it into Eq.~(\ref{uneq}), one obtains,
 \begin{eqnarray}
(k_{g\perp}-k_\perp)^2<\frac{x}{x_g} k_\perp^2
\label{kc}
\end{eqnarray}
which imposes an upper limit for the transverse momentum phase space integration that is unconstrained in the standard BFKL equation.  Such kinematic constraint and various different approximate forms have been extensively studied in the past~\cite{Ciafaloni:1987ur,Catani:1989yc,Andersson:1995ju,Kwiecinski:1996td,Kwiecinski:1997ee}.  Other NLL contributions to  in the small $x$ limit were discussed in Refs.~\cite{Iancu:2015vea,Iancu:2015joa,Lappi:2016fmu,Hatta:2016ujq,Zhou:2016tfe,Xiao:2017yya,Zhou:2018lfq,Zheng:2019zul}.

On the other hand, one can derive a different cutoff for phase space integration by taking into account the constraint from external kinematics.  From Eq.(~\ref{onshell}), one has,
 \begin{eqnarray}
 x = x_g+\frac{(k_{g\perp}-k_\perp)^2  }{2x_p P^+ P^-} \frac{\xi}{1-\xi}> x_g+\frac{(k_{g\perp}-k_\perp)^2  }{2(1-x_p) P^+ P^-} 
 \label{detx}
\end{eqnarray}
due to $\xi>x_p$. The lower limit for longitudinal momentum integration for real correction in Eq.~(\ref{sBFKL}) has to be correspondingly modified as,
 \begin{eqnarray}
\int_{x_g}^1 \frac{dx}{x} \  \longrightarrow \ \int_{x_g+\Delta x}^1 \frac{dx}{x}
\end{eqnarray}
where $\Delta x= \frac{(k_\perp-k_{g\perp})^2}{(1-x_p)s}$ is the minimal longitudinal momentum fraction carried by the emitted gluon.  This simple analysis obviously can be applied to the gluon emission from each rung of the BFKL ladder because the largest plus momentum component acquired by each emitted gluon must be smaller than $(1-x_p) P^+$.  Such kinematic constraint naturally leads to a  BFKL equation with a rapidity veto,
  \begin{eqnarray}
  G(x_g,k_{\perp}) \!\!&=&\!\! G_0(x_g,k_{\perp})   \label{int}
   \\ &+&\!\! \frac{ \alpha_s N_c}{\pi^2} \!\!  \int \!\! d^2 k_{g\perp}   \!\! \left \{ \int_{x_g+\Delta x}^{1}\!\! \frac{d x}{x} \frac{ G(x,k_{g\perp})}{(k_\perp-k_{g\perp})^2} 
 - \int_{x_g}^{1}\!\! \frac{d x}{x} \frac{k_\perp^2 G(x,k_{\perp})}{(k_\perp-k_{g\perp})^2 \left [k_{g\perp}^2+(k_\perp-k_{g\perp})^2\right ]  
 }   \right \}  \nonumber 
 \end{eqnarray}
Note that the virtual correction is not affected by the kinematic effect under consideration, and thus remains unchanged. 
 We now compare this new type kinematic constraint with the one that is extensively discussed in literature.  To this end, we re-express the on-shell condition  in the strong rapidity ordering region as,   
 \begin{eqnarray}
(k_{g\perp}-k_\perp)^2  =\frac{1-\xi}{\xi} \frac{x-x_g}{x_g} k_\perp^2< \frac{1-x_p}{x_p}\frac{x}{x_g} k_\perp^2
\end{eqnarray}
It is easy to see that at least in the last step of the evolution, this is a more stringent condition than that given by Eq.~(\ref{kc}) in the threshold region $x_p \rightarrow 1$, which is the main focus of the present work. Therefore, we will not impose the kinematic constraint  Eq.~(\ref{kc}), but only modify the lower  limit of $x$  integration for real correction part of the BFKL kernel as shown in Eq.~(\ref{int}).

Due to the limited available phase space for real gluon emissions,   the small $x $ evolution can not  fully develop as described by the standard LL BFKL equation.   One may anticipate that the kinematic constraint will significantly slow down   the small $x$  evolution, in particular, at large transverse momentum near threshold region. To investigate the impact of the NLL contribution from the kinematic constraint numerically, we derive the differential form of the BFKL equation, which can be straightforwardly obtained by  differentiating in $\ln \frac{1}{x_g}$ on both sides of Eq.~(\ref{int}),
\begin{eqnarray}
 \!\!\! \frac{\partial  G(x_g,k_{\perp})}{\partial \ln \frac{1}{x_g}} \!\!\!&=&\!\! \frac{\partial G_0(x_g,k_{\perp})}{\partial \ln 1/x_g}
  \\ &+&\frac{ \alpha_s N_c}{\pi^2} \!\! \int \!\! d^2 k_{g\perp} \!\! \left \{  \frac{x_g}{x_g\!+\Delta x}   \frac{  G(x_g\!+\Delta x,k_{g\perp})}{(k_\perp\!-k_{g\perp})^2}  -  \frac{k_\perp^2  G(x_g,k_{\perp})}{(k_\perp\!  -k_{g\perp})^2 \left [k_{g\perp}^2+(k_\perp\!-k_{g\perp})^2\right ]  
 }   \right \} \nonumber 
\end{eqnarray}
We use the MV model as the initial condition for the unevolved  gluon  TMD,
\begin{eqnarray}
 G_0(x_g,k_{\perp})=\frac{k_\perp^2N_c}{2\pi^2 \alpha_s }S_\perp \int \frac{d^2 r_\perp}{(2\pi)^2} e^{-ik_\perp \cdot r_\perp} \exp \left [-\frac{r_\perp^2 Q_{s0}^2}{4}\ln\left( \frac{1}{r_\perp \Lambda_{\text{mv}}}+e \right ) \right ]
\end{eqnarray}
where $S_\perp= 51$mb denotes the transverse area of the proton target, $Q_{0s}^2= \ 0.5 {\text{ GeV}}^2$ is the saturation scale at $x_0=0.01$.  $\Lambda_\text{mv}=0.241$ GeV is the infrared cutoff in the MV model. It is well known that the leading order BFKL gives a much too steep growth with $1/x$. This is to some extent related to the infrared diffusion of the BFKL dynamics, to avoid which, we numerically solve the kinematic constraint BFKL by imposing a saturation boundary, namely,  $\frac{\alpha_s(k_\perp) G(x_g,k_{\perp})}{k_\perp^2}=\frac{\alpha_s(k_\perp) G(x_g,k_\perp)}{k_\perp^2}|_{k_{\perp}=0.7 \text{ GeV}}$  when $k_\perp <0.7 \text{ GeV}$. Such phenomenological treatment for regularizing the infrared behavior is similar to that  proposed in Ref.~\cite{Avsar:2011ds}.

\begin{figure}[h!]\centering
\includegraphics[width=0.45\textwidth]{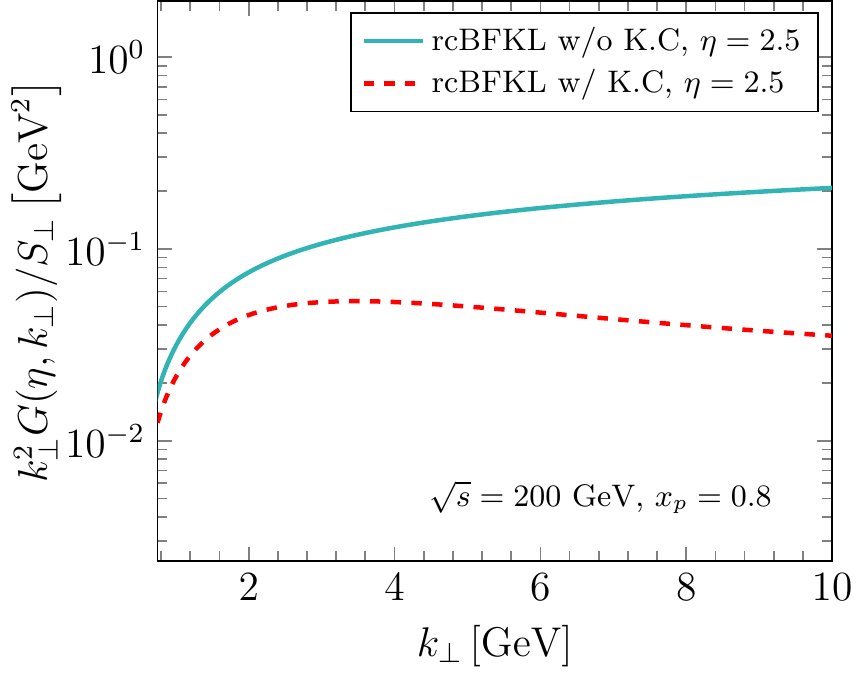}
\includegraphics[width=0.45\textwidth]{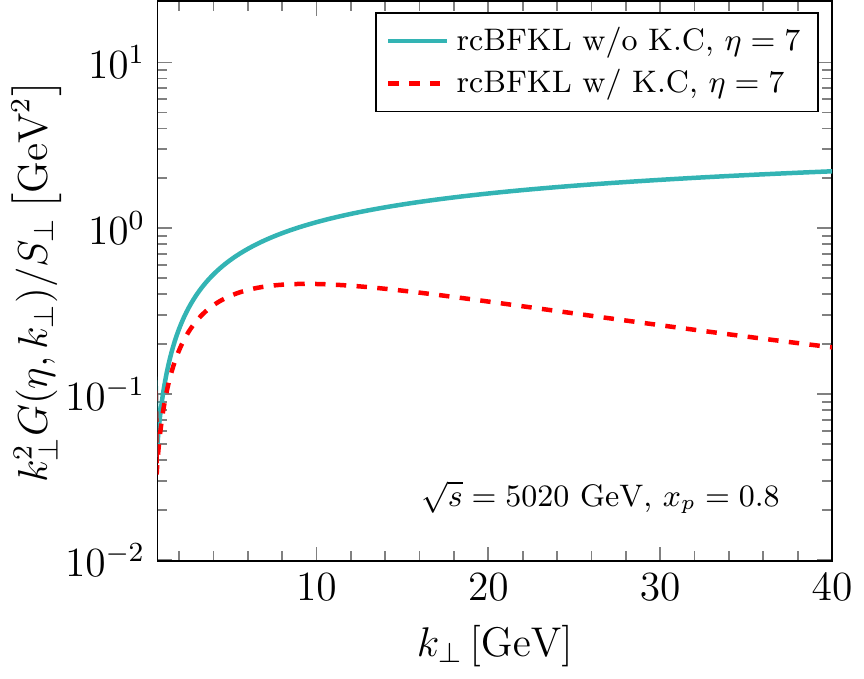}
\caption{The comparisons of gluon TMDs with and without the  kinematic constraint at RHIC and LHC energies with $\Delta x= \frac{(k_\perp-k_{g\perp})^2}{(1-x_p)s}$.
}
\label{kcbfkl}
\end{figure}

In the practical numerical implementations of the kinematic constraint BFKL evolution, we  used the running coupling prescription taken from Ref.~\cite{Kovchegov:2006wf}.
The numerical results  as a function of the $k_\perp$ for a given  $\Delta x$ are shown  in Fig.~\ref{kcbfkl}. As a comparison, the solution of the standard running coupling BFKL without kinematic constraint are also presented.  It is clear that the solution of the BFKL equation with a rapidity veto constraint gets suppressed  at large transverse momentum.

%

\section{Phenomenology  }
In this section,  we investigate the  impact of the kinematic constraint BFKL evolution on the phenomenology studies in two processes: inclusive forward particle production in pp collisions and  forward jet production in SIDIS processes. When the measured particle/jet carries the most longitudinal momentum fraction of the projectile, there is no much phase space left for the radiated gluons to drive the small $x$ evolution.  We shall observe that the particle/jet yield is  suppressed at relatively large transverse momentum due to the incomplete cancellation between the real and virtual corrections near threshold region.  

\subsection{Inclusive forward  particle production in pp  collisions }
Inclusive forward hadron production in pA/pp collisions is an important experimental observable to study dense gluonic matter in high energy scatterings~\cite{Dumitru:2001jn, Kharzeev:2003wz,Blaizot:2004wu,Blaizot:2004wv,Dumitru:2005gt,Albacete:2010bs,Guzey:2004zp,Altinoluk:2011qy}.  The suppression of particle yield at  low transverse momentum in the forward region has been argued to be one of the ``smoking gun" evidences of the saturation phenomena. Considerable experimental efforts have been made  in  identifying the saturation phenomenon via this observable at RHIC and LHC~\cite{BRAHMS:2004xry,STAR:2006dgg,ALICE:2018vhm,ATLAS:2016xpn,LHCb:2021vww}.
The complete NLO correction to the cross section  has been worked out in Refs.~\cite{Chirilli:2011km,Chirilli:2012jd}.

The numerical estimation~\cite{Stasto:2013cha}  however soon reveals that the NLO cross section of the inclusive forward  hadron production turns to negative at large  hadron transverse momentum.   
By taking into account the exact kinematics near the end point, the negativity problem can be partially remedied though the cross section still becomes negative at sufficient high transverse momentum~\cite{Watanabe:2015tja}.   A new factorization scheme~\cite{Iancu:2016vyg} using a running $x_g'$ in the hard part calculation also  offers a possible solution to this issue~\cite{Ducloue:2017mpb,Ducloue:2018lil}.  More recently, the reorganization of the perturbation series  based on the threshold resummation~\cite{Xiao:2018zxf,Liu:2020mpy,Shi:2021hwx} is shown to be a promising approach to maintain the positivity of the cross section as well.  Other attempts to fix this problem can be found in Refs.~\cite{Stasto:2014sea,Altinoluk:2014eka,Kang:2014lha,Ducloue:2016shw,Xiao:2014uba}. 

In this work, we adopt the subtraction scheme suggested in Ref.~\cite{Iancu:2016vyg}. However, in contrast to the original proposal~\cite{Iancu:2016vyg}, we subtract the modified BFKL kernel presented in Eq.~(\ref{int}) from the hard part instead of the standard one.   We show that the convergence of the perturbative series is greatly improved by absorbing the large threshold logarithm $\ln \frac{1}{1-x_p}$ arises in the NLO calculation  into the  kinematic constraint BFKL equation.

 For the demonstration purpose, here we only consider the quark initiated channel. The LO and the NLO cross section in the hybrid approach can be organized into the following form,
\begin{eqnarray}
\frac{d\sigma}{d^2p_{h\perp} dy}=\sum_f \int \frac{dz}{z^2} d \xi \left [  x_p q_f(x_p){\cal F}_{x_g}(k_\perp)   + H_s  +H_{ns} \right ]  D_{h/q}(z) 
\label{cs}
\end{eqnarray}
where the longitudinal momentum fractions carried by the incident quark and gluon are fixed by the external kinematics $x_p=k_\perp e^y/\sqrt s$, $x_g=k_\perp e^{-y}/\sqrt s$ with  $k_\perp$ being the transverse momentum transfer to quark from the target and $y$ being the rapidity of the produced hadron with the transverse momentum $p_{h\perp}=zk_\perp$. $q_f(x_p)$ and $D_{h/q}(z) $ are the quark PDF of the projectile proton and the collinear fragmentation function respectively. ${\cal F}_{x_g}(k_\perp)$ is the Fourier transform of the dipole amplitude which is related to the gluon TMD  via ${\cal F}_{x_g}(k_\perp)= \frac{2\pi^2 \alpha_s }{k_\perp^2N_c}G(x_g,k_\perp)$.

In order to be consistent with the implementation of the BFKL evolution, the NLO correction is  computed in the dilute limit following Ref.~\cite{Watanabe:2016gws}.  The NLO correction is separated into the singular part  $H_s$ and the non-singular part $H_{ns}$ with the help of the identity,
\begin{align}
    \int_\tau^1d\xi\frac{1+\xi^2}{1-\xi}f(\xi){\cal F}_{x_g'}=\int_\tau^1d\xi\frac{1+\xi^2}{(1-\xi)_+}f(\xi){\cal F}_{x_g'}+\int_0^1d\xi\frac{2}{1-\xi}f(1){\cal F}_{x_g'}
    \label{eq:plus-function}
    \end{align}
with  $f(\xi)$ being an arbitrary test function.  The longitudinal momentum fraction of the mother gluon is determined as $x_g'=x_g+x_g\frac{(k_\perp-k_{g\perp})^2}{k_\perp^2}\frac{\xi}{1-\xi}$ by external kinematics~\cite{Iancu:2016vyg}. Note that the "plus" prescription defined above is slightly different from the conventional one.    All contributions that are free from the rapidity  singularity at $\xi=1$, i.e. the first term on the right side of the above equation are grouped into $H_{ns}$ which reads,
\begin{eqnarray}
 && \!\!\!\!\! H_{ns}=\frac{\alpha_s}{2\pi^2}\frac{ x_p}{\xi} q_f(\frac{x_p}{\xi})\frac{1+\xi^2}{(1-\xi)_+}  \int d^2 k_{g\perp}  \left \{ \frac{ N_c}{k_{\perp}^2} {\cal F}_{x_g'}(k_{g\perp})\left[1  -\theta(k_\perp-k_{g\perp} ) -\theta(k_\perp-\xi k_{g\perp})  \right] \right .\ \nonumber \\ &&  \left .\ \!\!+  \frac{C_F }{(k_{g\perp}\!-k_\perp)^2}
  \Bigg[ {\cal F}_{x_g'}(k_{g\perp})-\frac{\mu^2 {\cal F}_{x_g}(k_\perp)}{(k_{g\perp}\!-k_\perp)^2+\mu^2} 
  +\frac{{\cal F}_{\tilde x_g'}\left(\frac{k_\perp}{\xi}\!+k_{g\perp}\!-k_\perp\right) }{\xi^2}-\frac{1}{\xi^2}\frac{\mu^2 {\cal F}_{x_g/\xi}\left(\frac{k_\perp}{\xi}\right)}{(k_{g\perp}\!-k_\perp)^2+\mu^2}  \Bigg]  \right \} \nonumber \\
  &&  
  +\frac{\alpha_s}{N_c2\pi^2}\frac{ x_p}{\xi} q_f(\frac{x_p}{\xi})\! \left[\frac{1+\xi^2}{(1-\xi)_+}\overline{I}^{(1)}_{qq}-\left(\!\frac{(1+\xi^2)\ln(1\!-\!\xi)^2}{1-\xi}\!\right)_+\!\! {\cal F}_{x_g'}(k_\perp)\right]\!\nonumber \\&&
  - \frac{N_c\alpha_s}{2\pi} x_p q_f(x_p){\cal F}_{x_g}(k_\perp)\int_0^1 \! d\xi^\prime\frac{1+\xi^{\prime2}}{(1-\xi^\prime)_+}\ln \xi^{\prime2}
  \end{eqnarray}
 where   $\tilde x_g'=x_g+x_g\frac{(k_\perp/\xi- k_{g\perp})^2}{k_\perp^2}\frac{\xi}{1-\xi}$.    $\mu$ is the factorization scale. We emphasize again that  the plus-prescription only acts on the $q_f$ while the $\xi$ dependencies of ${\cal F}$ and $\overline{I}^{(1)}_{qq}$ are unaffected.   The short hand notation
  $\overline{I}^{(1)}_{qq}$  stands for, 
\begin{eqnarray}
  \overline{I}^{(1)}_{qq}&=&\int d^2k_{g\perp}\left[{\cal F}_{x_g'}(k_{g\perp})\frac{(k_\perp-k_{g\perp})\cdot(k_{\perp}-\xi k_{g\perp})}{(k_\perp-k_{g\perp})^2(k_{\perp}-\xi k_{g\perp})^2} \right .\ \nonumber \\ &&\left .\
  -{\cal F}_{x_g'}(k_{\perp})\left\{\frac{(k_\perp-k_{g\perp})\cdot(\xi k_{\perp}-k_{g\perp})}{(k_\perp-k_{g\perp})^2(\xi k_{\perp}-k_{g\perp})^2}
  +\frac{k_{g\perp}\cdot(k_{\perp}-k_{g\perp})}{k_{g\perp}^2(k_{\perp}-k_{g\perp})^2}\right\}\right].
  \label{eq:Iqq}
  \end{eqnarray}
 In the above result,  the collinear divergences  have been removed by subtracting    the NLO PDF and the NLO integrated fragmentation function in the $\overline{\rm MS}$ scheme.  When making this subtraction, we replace  $1/\hat\epsilon$ with
$  \frac{1}{\hat \epsilon }=  \int \frac{ d^{2- \hat \epsilon } k_{g\perp}}{2\pi}  \frac{\mu^{2+\hat \epsilon} }{k_{g\perp}^2 (k_{g\perp}^2+\mu^2)}$ to simplify  the numerical estimations.

Now we turn to discuss the singular part. We proceed by first  imposing  the exact kinematic cut  as introduced in Ref.~\cite{Watanabe:2015tja}. The upper limit of $\xi$ integration is then modified as,
  \begin{align}
 \int_0^{1-\frac{(k_\perp-k_{g\perp})^2}{x_ps}}d\xi\frac{1}{1-\xi} =\ln \frac{1}{x_g}+\ln \frac{k_\perp^2}{(k_{g\perp}-k_\perp)^2}
 \label{rel}
 \end{align}
 For the virtual correction, the large logarithm $\ln \frac{1}{x_g}$ in the above formula is absorbed into the renormalized gluon TMD.  We are left with the second term which should be added back to the NLO hard part.  For the real correction,  the large logarithm arises in the first iteration of the BFKL evolution with a rapidity veto is  $\int_{x_g+\Delta x}^1 \frac{dx}{x} =\ln \frac{1}{x_g+\Delta x}$. After subtracting this small $x$ logarithm, one ends up with,
   \begin{align}
 \int_0^{1-\frac{(k_\perp-k_{g\perp})^2}{x_ps}}d\xi\frac{1}{1-\xi} -\int_{x_g+\Delta x}^{1} \frac{dx}{x}=\ln \frac{k_\perp^2+\frac{x_p}{1-x_p}(k_\perp-k_{g\perp})^2}{(k_\perp-k_{g\perp})^2}
 \label{vir}
 \end{align}
 which contributes to the finite part of the real correction.  At large $k_\perp$, the above expression is reduced to $\ln\frac{1}{1-x_p} $ which would be absent if one uses the LL BFKL equation. After taking into account the  the exact kinematic cut effect and subtracting the small $x$ logarithm, one arrives at the finite part, 
\begin{eqnarray}
 H_{s}&=&\frac{N_c\alpha_s}{\pi^2}  x_p q_f(x_p)\!\!\int \!\!d^2 k_{g\perp} \! \left \{ \int_{1-x_g-\Delta x}^{1-\frac{(k_\perp-k_{g\perp})^2}{x_ps}} \frac{d\xi }{1-\xi}{\cal F}_{x_g'}(k_{g\perp}) \frac{k_{g\perp}^2}{(k_\perp\!-k_{g\perp})^2k_\perp^2}   \right .\ \nonumber \\ && \left .\ - \int_{1-x_g}^{1-\frac{(k_\perp-k_{g\perp})^2}{x_ps}} \frac{d\xi }{1-\xi} {\cal F}_{x_g'}(k_{\perp}) \frac{k_{\perp}^2}{(k_\perp-k_{g\perp})^2(k_\perp^2+(k_\perp-k_{g\perp})^2)}   \right \}
\end{eqnarray}
One intriguing point which is worthy to be mentioned is that the above  kinematic cut has to  be imposed on both the real and virtual corrections simultaneously to ensure infrared finite, whereas the different treatments of kinematic constraint in Eq.~\ref{rel} and Eq.~\ref{vir} still lead to a infrared finite result.

Such subtraction scheme based on the NLL BFKL equation can be further refined by noticing that the probability for emitting a gluon within a given rapidity interval is not only determined by the integral $\int \frac{d\xi }{1-\xi}$, but also affected by the quark PDF $q_f(x_p/\xi)$. As well known, at large $x$, the quark PDF decreases  very quickly when $x$ approaches 1. To account for this effect, we introduce an effective $x_p'$ by solving the equation,
\begin{eqnarray}
\int_{x_p} d\xi \left [     q_f(x_p) -      q_f(\frac{x_p}{\xi}) \right] \frac{1}{1-\xi} =   q_f(x_p)  \ln \frac{1}{1-x_p'}
   \end{eqnarray}
where  $x_p'$ is always larger than $x_p$. Only if were quark PDF  an uniform distributed  in the range ($x_p$,1),  one has $x_p=x_p'$.  This procedure amounts to rearrange the contribution between the fixed order corrections and the NLL BFKL kernel.   To some extent, $x_p'$ plays a role similar to a running  factorization scale. In the following, we set $x_p'=0.8$ to fix the values of $\Delta x$ and $x_g'$  for simplifying the numerical calculations.

\begin{figure}[h!]\centering
\includegraphics[width=0.45\textwidth]{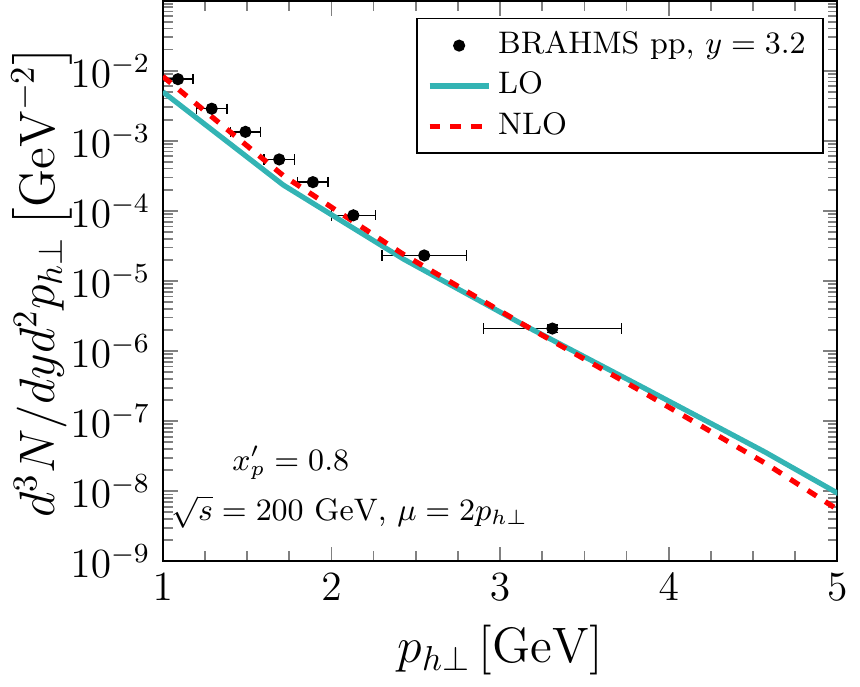}
\includegraphics[width=0.45\textwidth]{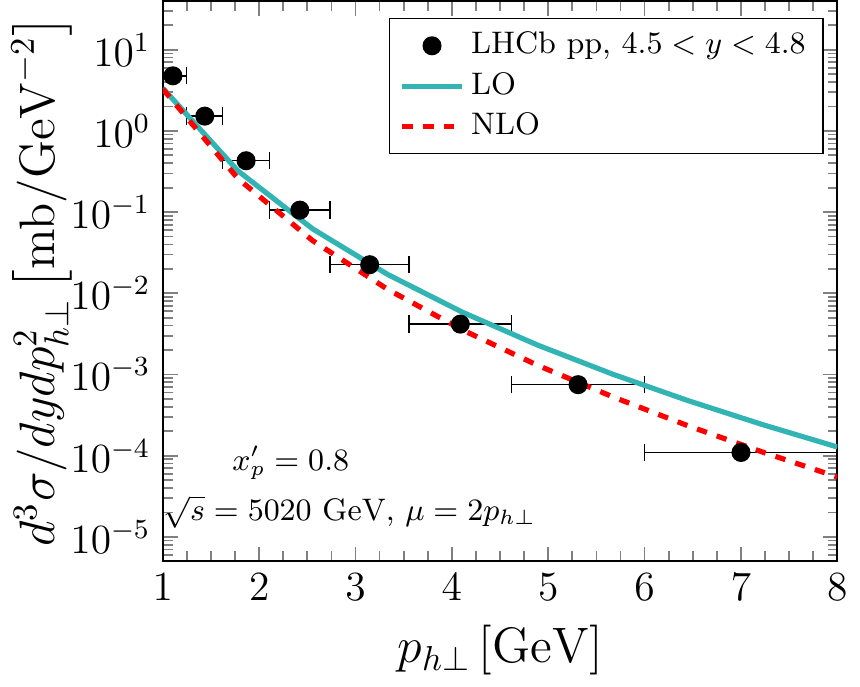}
\caption{Comparisons of the  NLO cross sections with data from RHIC ($\sqrt{s} = 200 $ GeV ) and LHC ($\sqrt{s} = 5020 $ GeV )   in pp collisions~\cite{BRAHMS:2004xry, LHCb:2021vww}. The LO results are obtained using the solution to the kinematic constraint BFKL equation.
}
\label{fig:lhcvsrhic}
\end{figure}

 We compute the inclusive forward hadron production in pp collisions at RHIC and LHC energies in the very forward region where the quark initiated channel  dominates. The  numerical results are presented in Fig.~\ref{fig:lhcvsrhic}.   One can see that the NLO result is smaller than the LO result almost in the entire kinematic range reached at LHC, which is line with the observation made in Refs.~\cite{Iancu:2016vyg,Ducloue:2017mpb,Ducloue:2018lil}, whereas the NLO contribution is larger than the LO one at relatively low transverse momentum for RHIC kinematics. These numerical results also confirm that the new subtraction scheme renders us to obtain a more stable NLO  contribution. This is because that  the part of sub-leading logarithm contribution at the NLO has been resummed by the kinematic constraint BFKL, and thus incorporated into the LO result. Moreover, the kinematic constraint BFKL appears to be a crucial ingredient for correctly describing the observed $k_\perp$ shape. Let us close this section with a final remark.  In order to fit experimental data,  we adopt a somewhat unrealistic large transverse area of proton:  $S_\perp= 51$mb following  Ref.~\cite{Shi:2021hwx}.  Our calculation would  underestimate the experimental measurements by the factor of two or three if we use the initial conditions for gluon distribution fitted to HERA data from a leading order calculation. It might be more appropriate to compute the NLO correction to the forward particle production using a fit from HERA data based on a NLO calculation~\cite{Beuf:2020dxl}, preferably with the kinematic constraint effect being taken into account.  As it is far beyond the scope of the current work, we  leave such NLO global fitting for the future study. 
 
\subsection{Forward jet production in SIDIS }
One of the golden channels for probing the saturation effect is the  di-jet production in the so-called correlation limit in semi-inclusive DIS process~\cite{Kovner:2001vi,Gelis:2002nn,Dominguez:2011wm,Metz:2011wb}. Due to the relatively large invariant mass of di-jet system that is required for reconstructing jets, it is however hard to access very small $x$ region at EIC through this observable. Moreover, the contribution to pair $k_\perp$ broadening from the Sudakov effect could also complicate the identification of  the saturation effects. Alternatively,  the saturation effect can be observed in the semi-inclusive production of a single hadron(or jet) as well~\cite{Mueller:1999wm}. In particular,  the sensitivity of this process to gluon saturation has been argued to be enhanced in the threshold region where the produced hadron(or jet)  carries a large longitudinal momentum fraction($z \approx 1$)  of the incoming virtual photon~\cite{Iancu:2020jch}.  This is precisely the region where the kinematic constraint discussed above could play an important role. 

In the leading order calculation of the SIDIS process, the minus component of  the exchanged gluons momenta is strictly set to be zero.  However, in reality,  the exchanged gluon carries a nonvanishing minus component of the light cone momentum acquired from the radiated gluons in the cascade  due to the recoil effect. Though this is formally a next leading power contribution, it can be very important in the threshold region as discussed in the previous section. To account for this effect, we explicitly keep the finite minus component of the exchanged gluons in the following analysis, and   denote the longitudinal momentum fractions of virtual photon shared by the produced jet, unobserved quark/anti-quark and the radiated gluon as $z$, $z_1$ and $z_2$ which satisfy the relation $z+z_1+z_2=1$ and $z_1\sim z_2 \ll z\sim 1$.

We only consider transverse photon production of jet,   as the longitudinal SIDIS cross section is suppressed by the factor $z_1$ in the threshold region. At the leading order, the transverse SIDIS cross section in the small $x$ formalism can be cast into a fairly compact form~\cite{Iancu:2020jch}, 
\begin{eqnarray}
\frac{d \sigma}{dz d^2 k_\perp}=2S_\perp N_c \frac{\alpha_{em}}{\pi} e_f^2\left [ z^2+(1-z)^2\right]
{\cal J}_T
\end{eqnarray}
with
\begin{eqnarray}
{\cal J}_T\!=\!\!
\int \! \frac{d^2r_\perp}{(2\pi)^2} e^{-i k_\perp \!\cdot r_\perp}\!
  \left \{ \frac{i k_\perp\! \cdot\!   r_\perp }{\bar Q |r_\perp| }  \frac{\bar Q^2K_1(\bar Q |r_\perp|)}{k_\perp^2+\bar Q^2} \!-\!\frac{1}{2} \!\left[\!  K_0(\bar Q |r_\perp|)-\!\frac{\bar Q |r_\perp|}{2
}K_1(\bar Q |r_\perp|)\right ] \right \}\! {\cal F}_{x_g}(r_\perp)
\end{eqnarray}
where $k_\perp$ is the produced jet transverse momentum. $K_0$ and $K_1$ are the modified Bessel functions. ${\cal F}_{x_g}(r_\perp)$ is the Fourier transform of ${\cal F}_{x_g}(k_\perp)$. $\bar Q$ is defined as $\bar Q^2=zz_1Q^2$ with $Q$ being the incoming photon's virtuality. The longitudinal momentum fraction of the probed gluons are determined by the external kinematics according to~\cite{Iancu:2020jch},
\begin{eqnarray}
x_g=\frac{1}{2 P\cdot q} \left (  \frac{k_\perp^2}{z}+ \frac{\text{max} (\bar Q^2, Q_s^2)}{z_1}+Q^2\right ) 
\end{eqnarray}
The above formula also can be easily converted into the momentum space expression~\cite{Dong:2018wsp}.

\begin{figure}[h!]\centering
\includegraphics[width=0.45\textwidth]{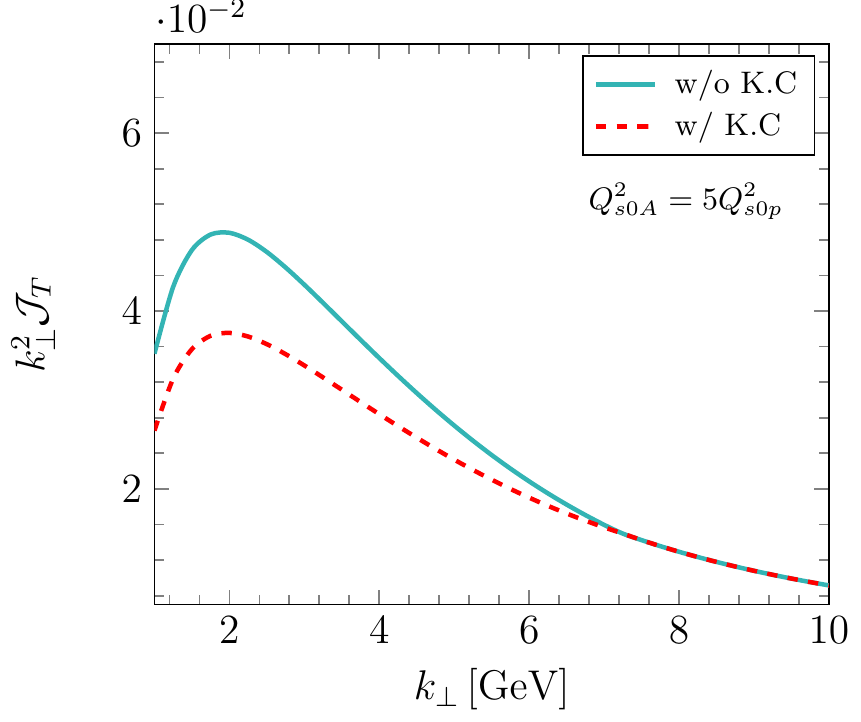}
\caption{The comparisons of  the transverse inelastic cross section multiplied by $k_\perp^2$ with and without the kinematic constraint.
}
\label{eic}
\end{figure}

Since the BFKL description  only can apply in the dilute limit, we restrict ourself to study the $k_\perp$ spectrum at large transverse momentum. The numerical results for the quantity $k_\perp^2 {\cal J}_T$ are presented as the function of $k_\perp$ in Fig.4.  At EIC energy, the kinematics are chosen to be $2 P\cdot q=10^4 \text{GeV}^2$ and $Q^2=10 \text{GeV}^2$.  The predication is made for the configuration  $z=0.8$, $z_1=0.1$ and $z_2=0.1$.
The numerical results are presented in Fig.~\ref{eic}. At  intermediate transverse momentum, the cross section is suppressed by roughly 20\%  due to the kinematic constraint effect. At high transverse momentum  where $x_g$ approaches $0.01$, the contribution  mainly comes from the initial gluon  distribution. Therefore, the results that are from the standard BFKL evolution and the kinematic constraint version converge  at high transverse momentum.

\section{Summary}

In this work, we have introduced a BFKL equation with the kinematic constraint near threshold region, which  approximately ensures  the longitudinal momentum conservation. Due to the rather limited phase space available for real radiation in the threshold region, such NLL BFKL generally slow down small $x$ evolution  in particular at high transverse momentum. It  differs from the conventional kinematic constraint BFKL equation results from requirement that the the offshellness of exchanged  gluon is dominated by transverse components.    We show that the subtraction of the kinematic constraint NLL BFKL kernel  from the unsubtracted hard part can  lead to a more stable NLO contribution to the inclusive forward hadron production in pp collisions  at high transverse momentum. We carry out  the detailed numerical studies of this observable for RHIC and LHC energies and found the good agreement with the experimental data.  We  investigated the impact of such NLL BFKL  evolution on the forward jet production in SIDIS process as well.

There are a number of directions in which the present work can be extended. First, a complete and sophisticated  phenomenological analysis of the observable would require us to take into account saturation effect using a running coupling BK equation with the same external  kinematic constraint. However, we do not expect that such NLL contribution arises from the kinematic effect plays an important role at low transverse momentum as it mainly modifies high $k_\perp$ tail behavior.   Second, one may notice that the longitudinal momentum conservation(along the projectile direction) is only approximately  kept in our treatment. To impose the longitudinal momentum conservation locally in the each step of the evolution,   it is  highly desirable to develop a Monte Carlo event generator based on the NLL BK equation. Lastly, it would be also interesting to apply our approach to   other processes, such as forward di-jet production in pA or eA collisions.

\section*{Acknowledgments}
We thank Shu-yi Wei for helpful discussions. 
Jian Zhou has been supported by the National Natural Science Foundations of China under Grant No.\ 12175118. Hao-yu Liu has been supported by the China Postdoctoral Science Foundation under Grant No.~212400211.

\bibliography{ref.bib}

\end{document}